# Imaging by means of a simple diffractive or refractive element: the axicon

José Joaquín Lunazzi and Daniel S. Ferreira Magalhães
Instituto de Física "Gleb Wataghin"
Universidade Estadual de Campinas, Unicamp
13083-970, Campinas, São Paulo, Brasil

In this paper[1] we demonstrate that the light diffracted by a simple compact disc can be used to generate images with basic attributes. We compare this attributes with the ones obtained with refractive elements. A compact disc is an axicon that generates a diffraction-free beam and we have shown that the focused position of the image depends on the wavelength of the diffracted light, thus it can be useful as a spectral filter. The experiments are of easy reproduction, allowing the understanding of images that the students observe daily at school or at home.

42.30.-d, 42.25.Fx, 42.30.Kq, 42.15.Eq

## Introduction

High quality photographic images have always been associated with high quality refractive objectives. This traditional view may limit the students' ability to think about imaging systems in a more general way, specifically in relation to the capability of transmission of light information even when the relationship between object and image is not straightforward. If an elaborate analysis is made, any light distribution arising from a luminous object point may have information about the object. A well corrected achromatic aplanatic objective is such an elaborate analog component. The comprehension and even perhaps the perception of new imaging systems might be visualized by students if their understanding is not limited to complicated and sophisticated refractive systems only. The knowledge of the properties of an optical element may be used to determine characteristics of an object under non-ideal imaging conditions. The history of optics shows that development of better imaging elements is a continuous evolution process. The introduction of diffractive systems as imaging systems may lead to new applications in future. As an example one can observe the development of a space telescope[2], where the primary lens is a large diffractive element.

According to the definition of H. McLeod *"an axicon is an optical element that images a point into a line segment along the optical axis"* [3,4,5]. It is known that an axicon is a non-conventional element, which is not useful for high quality images but it offers better depth of field than an usual set of lenses or objectives[6]. Because of its interesting properties, we have performed some experiments for a teaching laboratory. Before the term "axicon" was defined about fifty years ago, the axicon had generated many discussions that enhanced our knowledge of optics[7]. Traditionally, axicons are refractive and made of a glass cone. The basic properties of a refractive spherical or conical element can be easily demonstrated with domestic wine glasses. Diffractive elements may have the same properties, the focusing of a spherical lens can be represented by a Gabor zone plate and that of a conical lens by a circular element with constant period. A Gabor zone plate (GZP) has a geometrical distribution equivalent to the classical Fresnel zone plate but because of its sinusoidal transmission profile, it has only one convergent and one divergent focus. It can be constructed by photographic exposure of the interference of two coherent collinear light beams generating a holographic optical element (HOE). However a circular element with constant period can not be easily constructed by either optical interference, neither mechanically due to the large number of lines per millimeter required. The closest approximation can be found in a compact disc (CD), which is the optical element with constant period. Sochacki [3] made a theoretical and experimental comparison of the depth of focus between a computer-generated uniform-intensity axicon and a holographic Fresnel lens under monochromatic light and shown that though axicon image was not clear, it kept its properties over a much longer distances. Once detected, the presence of bulk or point objects can be observed over a longitudinal distance comparable to the focal length. In this paper we discuss the possibilities of a didactic application of a CD as an optical element that forms images under white light illumination.

## Description

The focal depth of refractive elements, such as can be demonstrated by simple experiments by spherical and conical elements ordinary wine glasses filled with water. The diffractive properties of transparent CDs are not well known. In order to facilitate the understanding of the imaging process proposed with this diffractive element, the properties of an ordinary CD, adapted to transmit light, is presented and demonstrated that it represented a conical element. We describe it as follows: the diffraction pattern produced by a compact disc (CD) under monochromatic illumination or by any kind of spiral structure was calculated by Ferrari [8]. The $n^{th}$ component of the field is

$$E_n(r',\theta',z) \cong n\, E_0\, c_{-n}\, \pi\, i^{-(n+1)} \exp(i\pi/4)\, (z\lambda/4r_0^2)^{1/2} \exp(i\, n\theta')\exp(-i\pi n^2 z\lambda/r_0^2) J_n(2\pi n r'/r_0) \quad (1)$$

for

$$(r_0/n\lambda)R_{min} < z < (r_0/n\lambda)R_{max} \quad (2)$$

Where r' and θ' are the coordinates of the field $E_n$ at the observation's plane; z is the distance between the CD and the observation's plane; $E_0$ is the incident collimated field; $r_0$ is the radial distance between adjacent turns on the CD; $c_{-n}$ is the nth component of a function that characterizes the profile of a CD (ex: a Bessel function of nth order); n is the diffraction order, in our case n=1; λ is the wavelength of monochromatic light; $R_{min}$ and $R_{max}$ are the minimum and maximum radius of the CD, respectively.

Outside the region given by eq. (2), this field is some orders of magnitude lower.

This analysis shows the formation of a diffraction-free beam [9] whose length depends on the wavelength, the period (spiral grooves) and the radial dimensions of the structure.

A special circular approach proposed by Magalhães [10], is easier to explain than the above spiral treatment. In many cases, the circular approach can explain images obtained with spiral structures - usually of dimensions much greater than the structure period. The scheme to understand the resulting pattern as a line is shown in Figure 1, under the following conditions: (a) illuminating of a small annulus of the disc with polychromatic light, (b) illumination of different regions of the disc with monochromatic light. For many point objects, the light pattern expected on the image field is a bundle of lines diverging from the center of the axicon that are sharply intersected at the imaging plane with a limited bandwidth along its longitudinal position - longer wavelengths corresponds to distances closer to the axicon; polychromatic imaging at some distance which is the center of the diffracting line; and shorter wavelengths standing at farther distances. At Figure 1(c) monochromatic light from all the CD forming two diffraction-free beam, n=1 and n=3 (all the other diffraction orders are presents too). Figure 1(d) exemplifies a very common situation everybody is acquainted when a small light source impinges a disc: a color line emerging from a CD or DVD like a holographic image, in full parallax and realm.

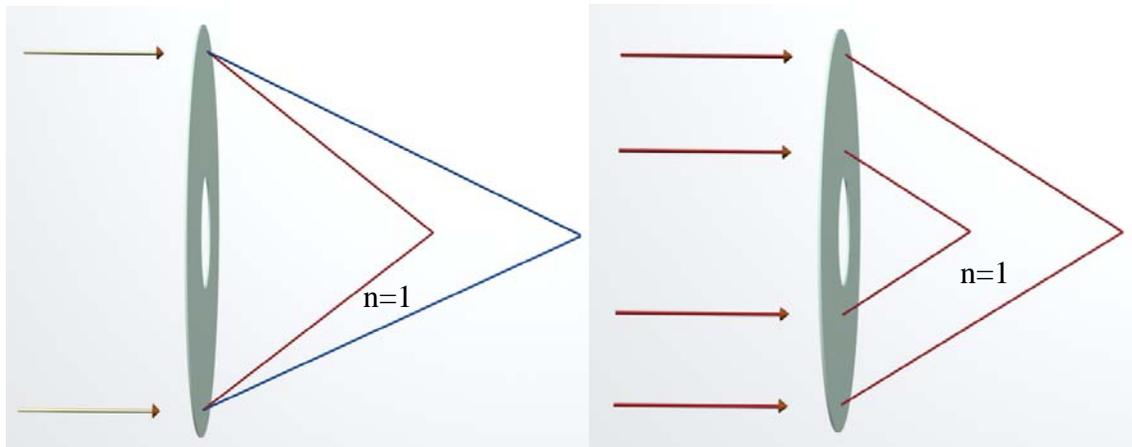

(a)                                    (b)

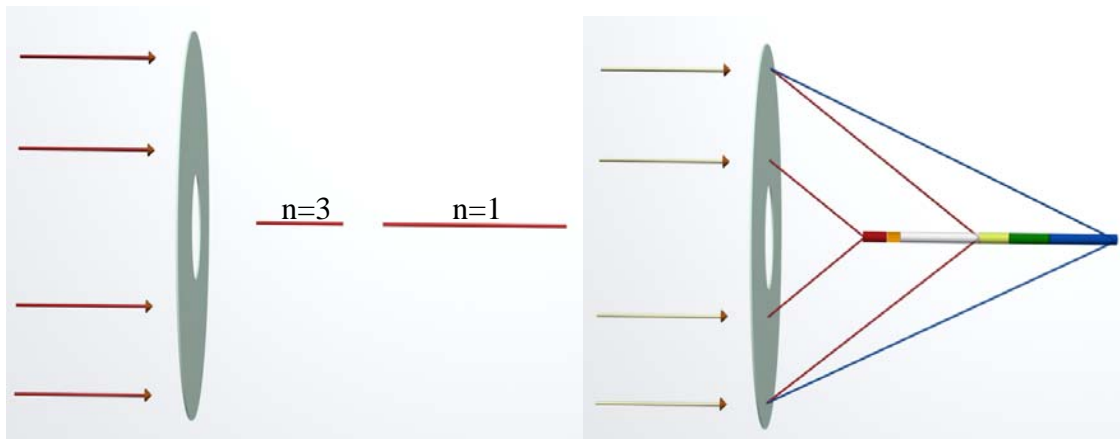

(c)                                    (d)

Fig 1: (a): polychromatic light from a small annulus of the compact disc (first diffraction order, n=1). (b): monochromatic light from different regions of the disc (first diffraction order, n=1). (c): monochromatic light from all the CD forming two diffraction-free beam, n=1 and n=3 (all the other diffraction orders are presents too). (d): polychromatic light forming an emerging color line from a CD (all the diffraction orders are presents)

## Experimental Setup

Some experiments are initially performed to compare the properties of a spherical vs. a conical lens. Using simple elements we obtained images which although strongly aberrated, can be interpreted qualitatively in terms of their aberrations to get information on the object.

In the first experiment, an almost spherical wine glass with water is used, Figure 2(a). Six objects are shown in Figure 2(b) and its off axis image is observed in Figure 2(c). We can observe that point objects create images, which are not points but due to the optical imaging element, while diffuse objects create diffuse but related images.

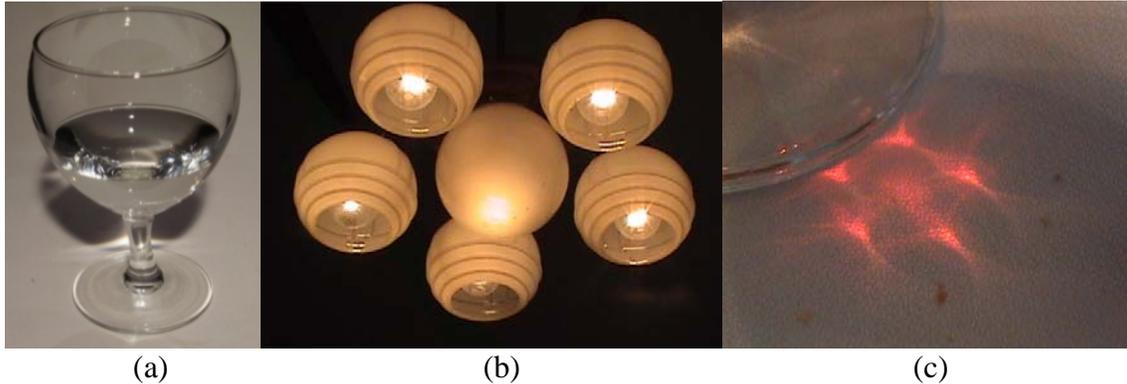

(a)                        (b)                        (c)
Fig 2: (a) An almost spherical wine glass with water. (b) Six objects. (c) The off axis image of the objects.

To demonstrate the axicon properties, a second experiment was performed to show the limited depth of field of a spherical refractive element: In Figure 3(a), a lamp bulb filled with water represents a spherical plano-convex lens. The radius and thickness of the lens were (3.25 ± 0.05) cm and (2.8 ± 0.1) cm, respectively. In Figure 3(b), the image of a point source at (112±1) cm distance is observed on the off-axis focus at (9.8 ± 0.1) cm from the bulb. Figure 3(c) shows the corresponding image at distance (26.3±0.1) cm. While traveling this short distance, the light from the object has spread so much that practically no image could be observed – thus it cannot be considered a good image of the point object.

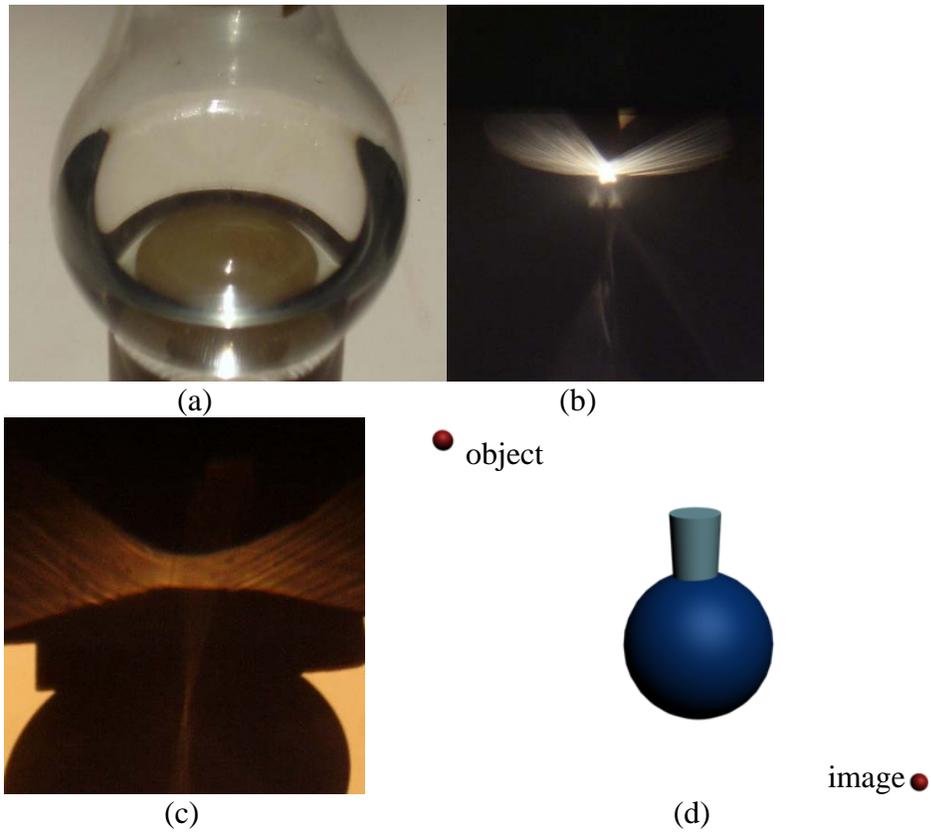

(a) (b)

object

(c) (d)

image

Fig 3: (a) A lamp bulb with water representing a spherical plano-convex lens. (b) The image of a point source in off-axis focus. (c) The spread image of the point source. (d) Geometrical situation of the imaging process.

The third experiment shows formation of an axicon in Figure 4(a) by a conical wine glass with water. In Figure 4(b), the off-axis image of a point object at (9.3±0.2) cm from the wine glass is shown. In Figure 4(c), the image at (34.3 ± 0.2) cm shows that the focused point image still holds the characteristics of a point, thus, it is a good image of the point object.

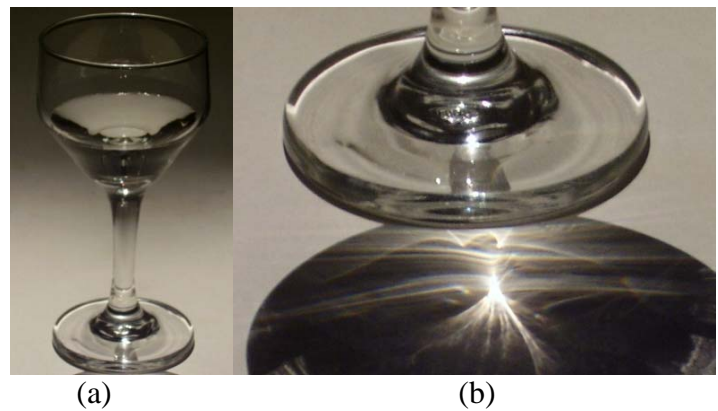

(a) (b)

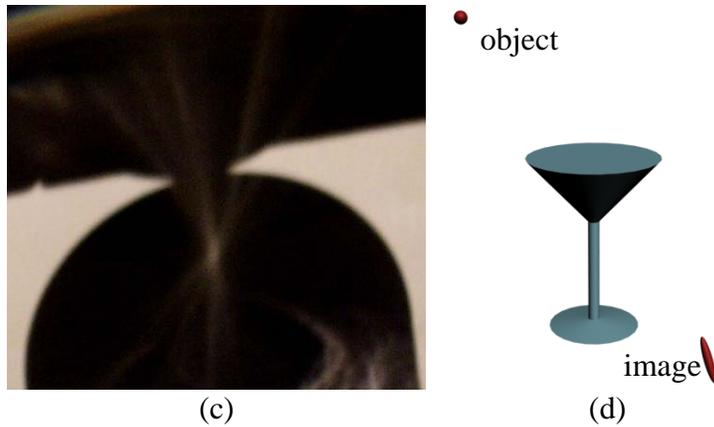

(c)                      (d)

Fig 4: (a) A conical wine glass with water. (b) The off-axis image at (9.3±0.2) cm of the point source. (c) The image at (34.3±0.2) cm keeps the characteristics of the point source. (d) Geometrical situation of the imaging process.

The fourth experiment employed a diffractive axicon and its optical action was characterized by observing the line it generates under illumination of monochromatic red light. A laser pointer without its collimating lens was located at (100±1) cm and used as a diverging point source. The image of the laser point was a line starting at (5.0±0.5) cm and ending at (13.0±0.5) cm, as shown in Figure 5.

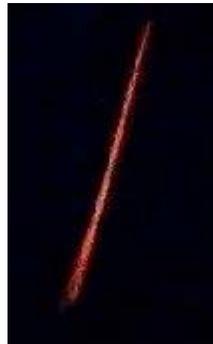

Fig 5: Diffraction-free beam length for the red light laser.

The fifth experiment employed a diffractive axicon similar to the previous experiment and a photographic camera was positioned on the imaged line to record the observation at this position. The light from a small fluorescent lamp bulb passing through a circular hole of 2 mm was used as a light source. Rings of different colors corresponding to the spectral lines of the source are observed and position of spectral rings cover different regions of the disc, which depend upon the position of the camera[11] (Figure 6).

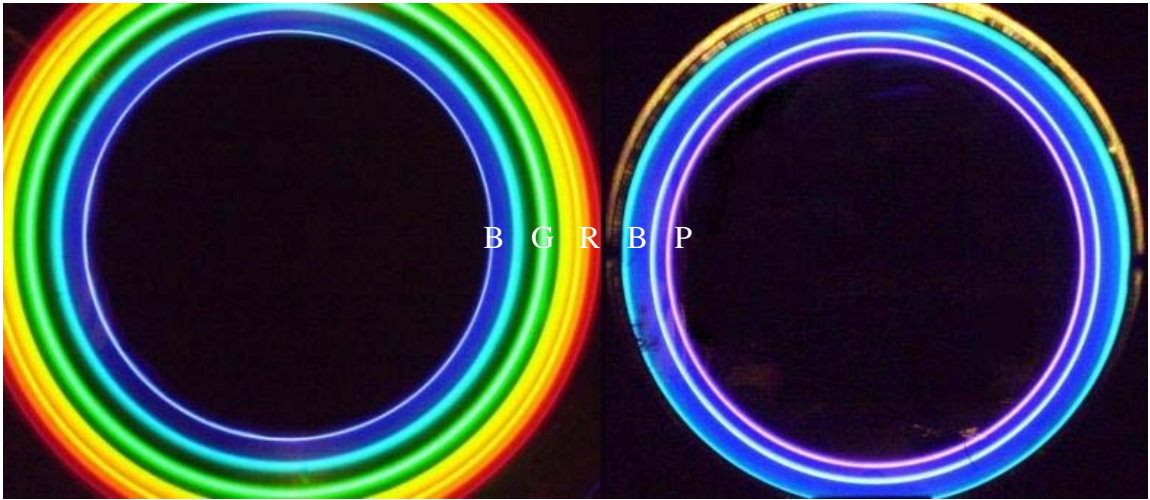

                (a)                                 (b)

Fig. 6: (a) The camera at position z=(9.2±0.1) cm, the spectral lines of the source B (blue), G (green) and R (red) are observed. (b) The camera at position z=(21.5±0.1) cm, the spectral lines of the source B (blue) and P (purple) are observed.

The sixth experiment was the photography of a Christmas tree with a string of a hundred small colorful light bulbs (Christmas candles of 0.5W each) placed within a distance interval $\overline{Z_0}$ between 4.2 to 4.4 m from the CD, as shown in Figure 7. The image was recorded on a color negative film Pro image 100 Kodak Professional (F), which was used in a Yashica FX-3 camera body. The lamps were distributed by color along a height of 1m.

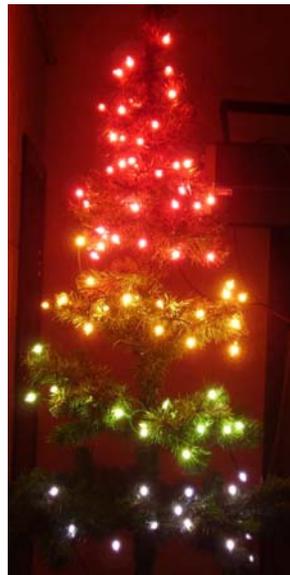

Fig 7: The object. A string of a hundred small colorful light bulbs (of power 0.5W each)

Displacing the film with respect to the CD-disk along the z-axis, we could see the image formation of each color-lamp set occurring inside a length interval. The images of the set of lamps at different positions were recorded with exposure time of $(1.0 \pm 0.1)$ s.

An ordinary compact disc without the reflective layer was used to work in transmission light in order to avoid shadowing of the image by the photographic film fixed to a support. The experiment is shown in Figure 8. Compact discs without the reflective layer are found as separators in most cylindrical CD-R boxes for fifty or hundred units. Otherwise the layer can be removed from compact discs by cutting the edges and dismounting its sandwiched composition. The central hole of the CD was covered to stop the light passing through it. The diffraction efficiency of the CD was estimated by measuring intensities of the first diffraction order of an ordinary red laser pointer and the transmitted intensity and it was found to be $(11.1 \pm 0.8)$ %. The spatial frequency of the CD ($\nu$) was calculated by using the diffraction equation of a grating (eq. 3).

$$\nu = \frac{sin\theta_i + sin\theta_d}{\lambda} \qquad (3)$$

Where $\theta_i$ is the incident angle; $\theta_d$ is the diffracted angle; $\nu$ is the spatial frequency and $\lambda$ is the wavelength of the light. The spatial frequency obtained was $670\pm90$ lines/mm.

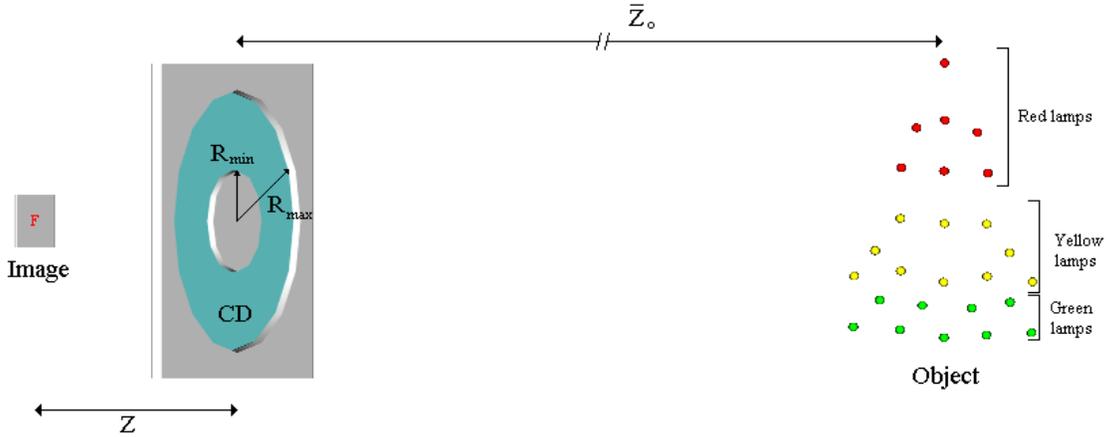

Fig 8: Experimental setup for image capture. Z is the distance between the film F and the CD. $\overline{Z}_0$ is the distance between the object and the CD. $R_{min}$ and $R_{max}$ are the minimum and maximum radius of the CD.

The image of the set of lamps recorded on the film at different positions on the z-axis is shown in Figures 9, 10 and 11. The figures are in the same size ratio, but inverted (top-bottom) to facilitate the comparison with the real object (Figure 7). In Figure 9(a), it is observed that the red light (lamps surrounded and labeled as R) is more intensive in the beginning of the image formation. In 9(b), we can see the red, yellow and green light bulbs (labeled as R, Y and G). As the z-distance was increased up to 12.5 cm (Figure 10(a)), no big change in the color distribution of each lamp was observed. In Figure 10(b), a change in the image of the red lamps in form of change of colors from red to violet is perceived. This color change caused intensity reduction in the black-and-white photo. A change of colors also occurred from yellow to violet, which is shown in Figures 11(a) and 11(b).

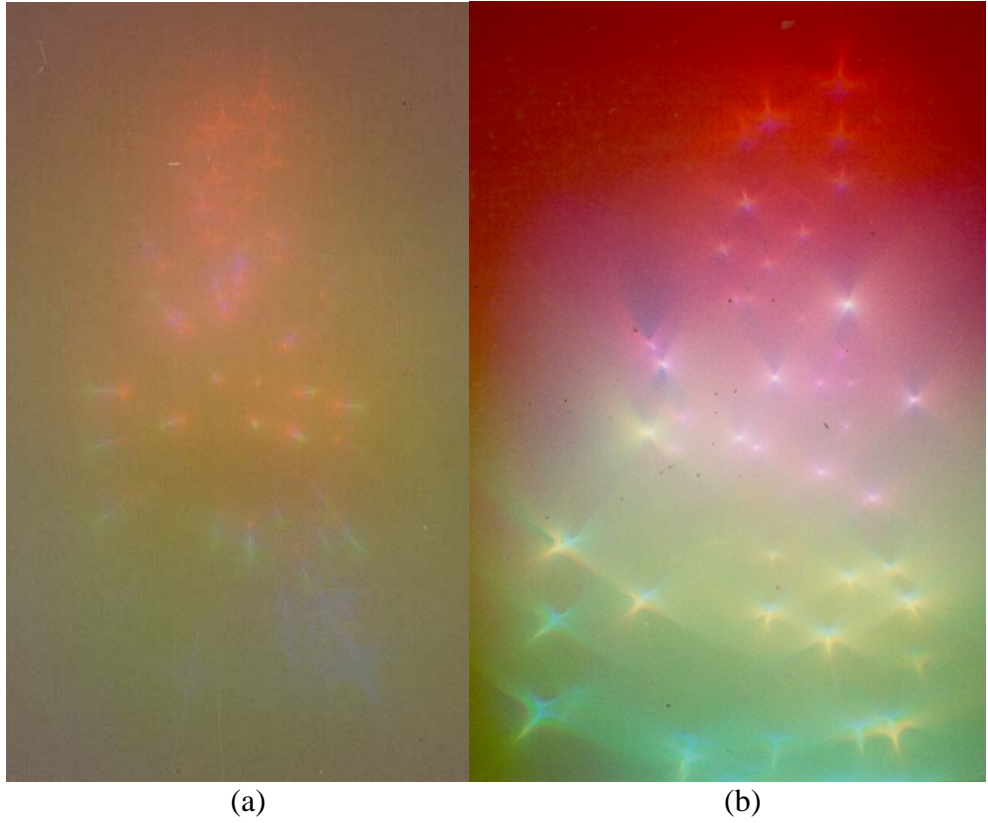
(a) (b)
Fig 9: Images of the lamps at Z distances: (a) (4.5±0.3) cm and (b) (9.0±0.3) cm

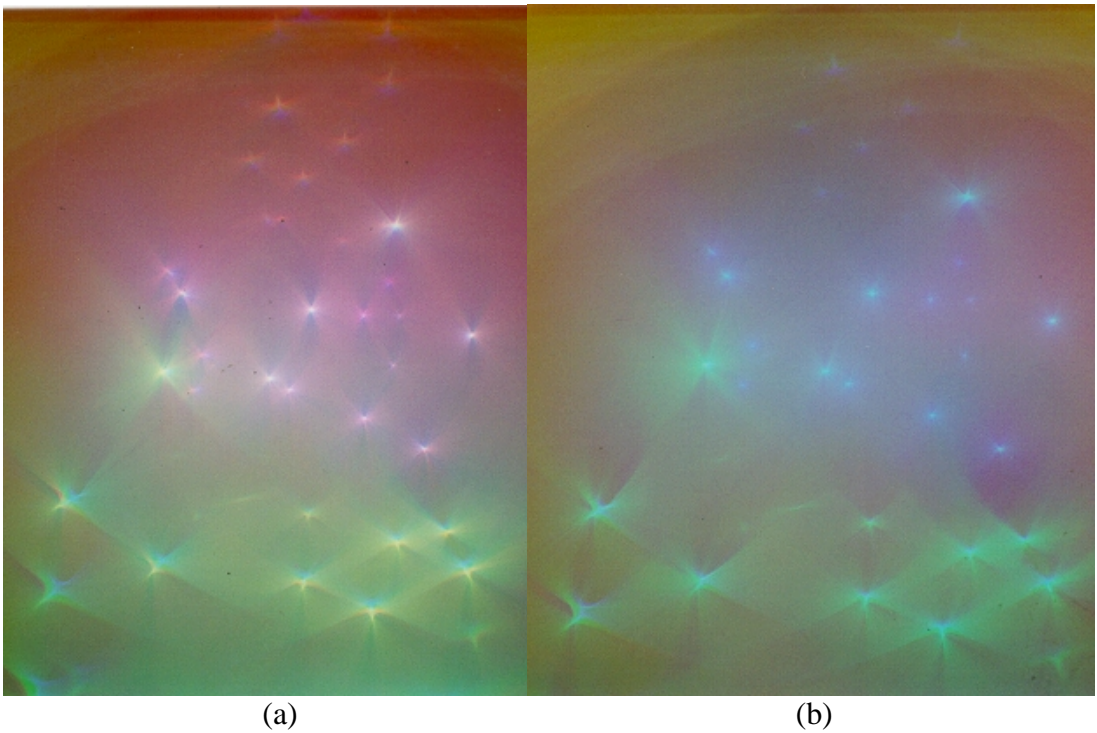
(a) (b)
Fig 10: Images of the lamps at Z distances: (a) (12.5±0.3) cm and (b) (14.0±0.3) cm

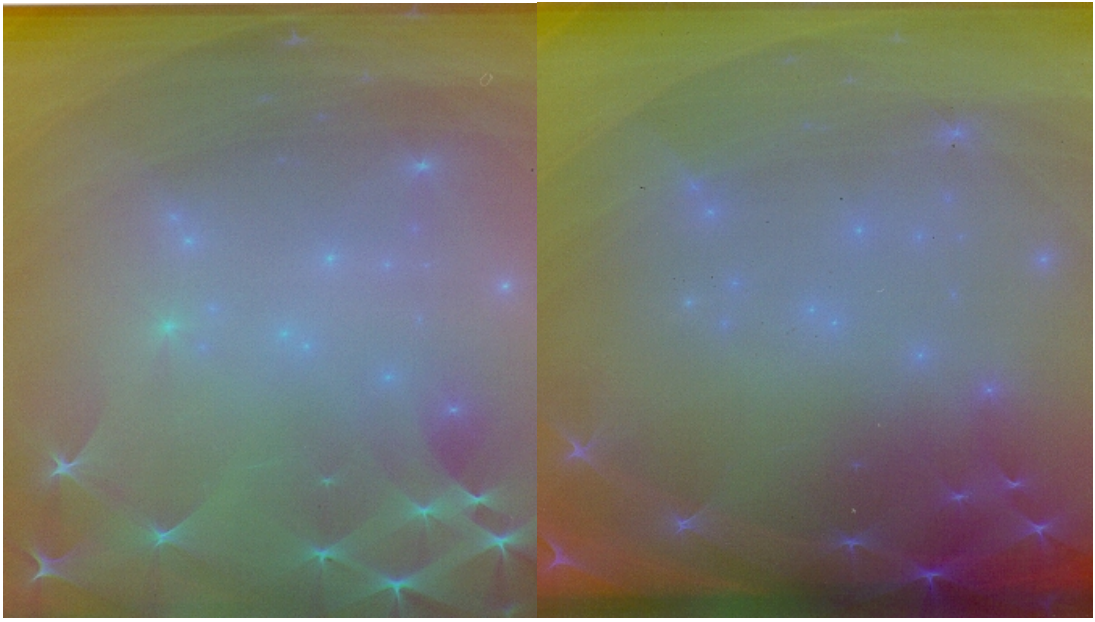
(a) (b)

Fig 11: Images of the lamps at Z distances: (a) (16.0±0.3) cm and (b) (16.5±0.3) cm

## Discussions

Using a mean wavelength and its deviation from its mean value[12], the depth of focus is studied for the colors of the objects. By means of Ferrari [8] and Magalhães [10] results shown in Equation 2, we could compare the laboratory results with calculated ones (Table I).

|  | Theoretical Results | | Experimental Results | |
| --- | --- | --- | --- | --- |
| Color | $Z_{min}$ (cm) | $Z_{max}$ (cm) | $Z_{max}$ (cm) | Results agreement |
| Red ($\lambda_R$=680±60 nm) | 4.8±0.6 | 13±1 | 13±1 | 100% |
| Yellow ($\lambda_Y$=580±10 nm) | 5.7±0.5 | 15±1 | 16.3±0.6 | 91% |
| Green ($\lambda_G$=530±30 nm) | 6.2±0.6 | 16±2 | - | - |
| Violet ($\lambda_V$=410±30 nm) | 8.0±0.9 | 21±2 | - | - |

Table I: Results of the diffraction-free beams length with $R_{min}$=(2.2±0.1)cm, $R_{max}$=(5.8±0.1)cm and $r_0$=(1.5±0.1)μm. The error in Z was calculated by Error Propagation [13].

The depth of focus of the system is directly related with the length of the diffraction-free beams. For white light objects, this depth begins at the red $Z_{min}$ and terminates at the violet $Z_{max}$. The length of the depth of focus is 16cm, Table 1. We could determine experimentally that the depth of focus was much greater than that corresponding to a photographic objective of the same focal length and aperture ($90\pm4$ cm$^2$). By replacing the light bulbs by purer color sources like LEDs enables to observe the vanishing of the images at the end of the respective diffraction line. The path difference *p* (see Dyson [14]) is given by:

$$p = \tfrac{1}{2} \rho^2 \theta^2 \cos^2 \phi [\tfrac{1}{Z_0} + \tfrac{1}{z}] \tag{4}$$

where $\rho$ and $\phi$ are the polar coordinates of a point in the plane of the spiral diffraction grating; θ is the incident angle on the spiral diffraction grating; $\overline{Z_0}$ is the distance between the object and grating and z is the distance between grating and image.

The absence of a term in cos $\phi$ indicates that there is no aberration of a comatic nature and the term in $\cos^2\phi$ represents astigmatism, that is the same in magnitude as the astigmatism of a simple lens. Unlike conventional astigmatic images, the two degenerated focal lines can be seen at once, giving a characteristic cross-shaped appearance to the image of a small object. This astigmatism becomes significant for an incident angle of about 7º.

The luminous efficiency of the system, restricted by the efficiency of the diffraction element, is about 11%. A highly efficient (almost 100%) diffraction grating, reported by Marciante [15][16], generates expectation of evolution in this area leading to development of practical systems possible in the future.

## Conclusions
By means of simple elements we obtained images with longer depth of focus than the corresponding images obtained with a spherical lens, and observed spot size evolution for spherical and conical elements. We show that it is possible to get diffractive images of point objects and to use them with the advantage of spectral separation along a longitudinal field with a large depth of focus. The efficiency of the system is restricted by the diffraction efficiency of the diffraction element and by astigmatism, which is present for large angles of incidence. The simplicity of the system makes possible the observation and verification of an unusual phenomenon in a regular classroom.

## Acknowledgements
The authors thank the "Coordenação de Aperfeiçoamento de Pessoal de Nível Superior" CAPES of the Brazilian Ministry of Education and the "Comissão de Pós Graduação" CPG of the Institute of Physics -Unicamp for financial support. The authors also thank J. A. Campos and T. Dellariva for technical support. Special thanks to Dr. Susana Souza Barros and one of the reviewers for English language support.